\documentclass[fleqn,twoside]{article}

\usepackage[headings]{espcrc2}
\usepackage{graphicx,epsf}
\readRCS
$Id: espcrc2.tex,v 1.2 2004/02/24 11:22:11 spepping Exp $
\usepackage{graphicx}
\usepackage[figuresright]{rotating}

\newcommand{\AmS}{{\protect\the\textfont2
  A\kern-.1667em\lower.5ex\hbox{M}\kern-.125emS}}

\newcommand{\bff}[1]{\mbox{\boldmath ${#1}$}}

\newcommand{\MS}{$\overline{\mathrm{MS}}$\ }

\newcommand{\be}{\begin{equation}}
\newcommand{\ee}{\end{equation}}
\newcommand{\bea}{\begin{eqnarray}}
\newcommand{\eea}{\end{eqnarray}}

\title{Third order Coulomb correction to $t\bar{t}$ threshold cross
  section \thanks{PITHA 06/03. Talk based on
  Ref.\cite{beneke-kiyo-schuller05}. To
appear in the proceedings of the 7th International Symposium on
Radiative Corrections (RADCOR05), Shonan Village, Japan Oct. 2005}}

\author{Y. Kiyo \thanks{
Supported by the DFG Sonderforschungsbereich/Transregio 9
``Computer-gest\"utzte Theoretische Teilchenphysik''.
}
\\[0.3cm]
{\sl Institut f\"ur Theoretische Physik E, RWTH Aachen,\\
D -- 52056 Aachen, Germany}}

\runtitle{Third order Coulomb corrections}
\runauthor{Y. Kiyo}

\begin{document}

\begin{abstract}
We report on our result of third order Coulomb correction to the 
cross section $\sigma(e^+ e^-\rightarrow t\bar{t})$ near threshold.
Analytic expression for the Coulomb energy and wave function at the
origin are obtained. We discuss the significance of the Coulomb
correction to the threshold cross section and heavy quarkonium
phenomenology. \vspace{1pc}
\end{abstract}

\maketitle

\section{Threshold cross section}
The $t\bar{t}$ threshold cross section has the following schematic
form in the conventional perturbative expansion of $\alpha_s$,
\begin{eqnarray}
\sigma(e^+e^-\rightarrow t\bar{t}) \sim \sigma_{\rm Born}
\large[1+\frac{\alpha_s}{v}+ \left(\frac{\alpha_s}{v}\right)^2 +
\cdots \,\large]\, ,
\end{eqnarray}
where $\sigma_{\rm Born}$ is a Born cross section, and
$v=\sqrt{1-4m_t^2/s}$ is speed of produced $t,\bar{t}$. The
combination of $(\alpha_s/v)^n$ appears to all order in the
perturbation theory, known as Coulomb singularity. The
perturbative expansion in $\alpha_s$ is not applicable to the
threshold cross section because $v$ is of order $\alpha_s$ near
the $t\bar{t}$ threshold $\sqrt{s}\sim 2m_t$, and the Coulomb
singularity $(\alpha_s/v)^n$ dominates the cross section. To
obtain meaningful cross section the Coulomb singularity has
to be summed to all order in $\alpha_s$. The physical origin
of the Coulomb singularity is instantaneous gluon exchange
between $t$ and $\bar{t}$ which has a small spatial momentum
$|\vec{q}|\sim v \sim 0$. This is called potential gluon because of its
propagator
\begin{equation}
\widetilde{V}_C(\bff{q}) = -\frac{4\pi C_F\alpha_s}{\bff{q}^2} \, ,
\label{Cpot}
\end{equation}
where $C_F=4/3$, which is the Coulomb potential $V_C =-C_F\alpha_s/r$
in coordinate space.

This understanding leads to a quantum mechanical
description which is equivalent to resummation 
of the Coulomb singularity. People had been looking for 
an effective field theory (EFT) which makes the resummation 
systematic. Finally a non-relativistic version of 
QCD was derived, called pNRQCD/vNRQCD 
\cite{Pineda-Soto97,Luke-Manohar-Rothstein99}. 
The EFT makes the resummation systematic based on 
nonrelativistic power counting 
$v \sim \alpha_s\ll 1$ in Lagrangian level. 
Nowadays we understand the resummation using the EFT, and
higher order corrections are taken into account in 
the EFT framework systematically. The NNLO calculation for 
the total cross section was completed by several groups 
\cite{TopWGR}, and now we are working on the NNNLO corrections
using the EFT.

The EFT classifies the corrections into
three classes:
\\
\\
$\bullet$ Hard corrections included in Wilson coefficients of
(composite)
operators.\\
$\bullet$ Potential corrections, which are non-local (in space)
4-Fermi operators but local in time. \\
$\bullet$ Dynamical gluon corrections called ultra-soft gluon in
the EFT.
\\
\\
The hard corrections are related to re-normalization of the
operators in the EFT. The ultrasoft correction
appears at NNNLO calculation, and the corresponding energy level
correction is know by Kniehl-Penin \cite{kniehl-penin99}. However
complete NNNLO correction to the total cross section is not known,
yet. We discuss a part of the potential corrections in this report, 
which has the following form in the EFT Lagrangian,
\begin{eqnarray}
&& \hspace{-.6cm} \delta {\cal L}(x)=\int d^3{\bff{r}}
\big[\psi^\dag \psi \big](x+\bff{r}) V(\bff{r}) \big[
\chi^\dag\chi\big](x)\, ,
\label{EFTLag}
\end{eqnarray}
where $\psi^\dag, \chi$ is a creation operator of heavy quark and
anti-quark, respectively. We will report on the result of the
Coulomb correction.

We parameterize the momentum space Coulomb potential as
follows
\begin{eqnarray}
\widetilde{V}(\bff{q}) = \widetilde{V}_C(\bff{q})\,
 \big[1+ \widetilde{\cal V}_{1}(\bff{q})
 +\widetilde{\cal V}_{2}(\bff{q})
 +\widetilde{\cal V}_{3}(\bff{q}) \big], \label{VCparametrization}
\end{eqnarray}
where ${\cal V}_{n}(\bff{q})$ is n-th order correction to the pure
Coulomb potential $\widetilde{V}_C$, induced by loop diagrams when
the EFT is derived from QCD. Explicit form of the potentials were 
derived at NNNLO (except $a_3$) in ref. \cite{kniehl-penin-smirnov-steinhauser00}, 
and the Coulomb potential reads
\begin{eqnarray}
&&\hspace{-.7cm} \widetilde{\cal V}_1 = \frac{\alpha_s}{4\pi}
\big[a_1+\beta_0 \, l_q \, \big]
\nonumber \\
&& \hspace{-.7cm} \widetilde{\cal V}_2 =
\left(\frac{\alpha_s}{4\pi}\right)^{\!2} \big[a_2 + \big(2
a_1\beta_0+\beta_1\big) \, l_q + \beta_0^2 \,l_q^2 \,\big]
\nonumber \\
&&\hspace{-.7cm} \widetilde{\cal V}_3 =
\left(\frac{\alpha_s}{4\pi}\right)^{\!3} \big[a_3 + 8\pi^2
C_A^3\,l_q(\nu) + \big(3 a_2\beta_0+2 a_1\beta_1
\nonumber\\
&&\hspace{-.5cm} +\beta_2\big)\,l_q + \,\big(3
a_1\beta_0^2+\frac{5}{2}\beta_0\beta_1\big) \, l_q^2 + \beta_0^3 \,
l_q^3 \, \big] ,
\label{CalV}
\end{eqnarray}
where $l_q=\ln(\mu^2/\bff{q}^2)$ and
$l_q(\nu)=\ln(\nu^2/\bff{q}^2)$, $\beta_i$ are the coefficients of
QCD $\beta$-function. The scale $\mu$ is QCD renormalization scale,
and the $\nu$ in ${\cal V}_3$ is a scale introduced  to separate 
the ultrasoft and potential modes in the EFT. Physical
quantities are scale independent if all the corrections at given order
are taken into account (see for instance
\cite{brambilla-pineda-soto-vairo99}). Now our task is to calculate
the threshold cross section using the EFT Lagrangian eq.(\ref{EFTLag})
with the Coulomb potential eqs.(\ref{VCparametrization}),
(\ref{CalV}).

\section{Green function method}
We use the optical theorem to calculate the threshold total cross
section which tells us that the cross section can be obtained
by taking the imaginary part of the correlation function of production 
currents
\begin{eqnarray}
&& \hspace{-0.6cm} \sigma(e^+ e^- \rightarrow t\bar{t}) =\frac{18\pi
e_t^2}{m_t^2}
{\rm Im} G(E+i \Gamma_t), \nonumber \\
&& \hspace{-0.6cm} G=\frac{i}{N_c}\int d^4 x e^{iqx}\langle {\rm
vac}|[\psi^\dag\chi](x) [\chi^\dag \psi](0)|{\rm vac}\rangle ,
\end{eqnarray}
where $N_c=3$ and $m_t$ is the quark pole mass. Using the EFT one
can show that the matrix element $G(E)$ can be expressed by the 
quantum mechanical Green function
\begin{eqnarray}
&& \hspace{-.7cm} G(E)=\langle 0| \hat{G}(E) |0\rangle =\langle 0|\,
\frac{1}{\frac{\bff{p}\,^2}{m_t}+V(\bff{r})-E}\,|0\rangle\, ,
\label{GreenFunc}
\end{eqnarray}
where $|0\rangle$ denotes a quantum mechanical position eigenstate
at the origin $\bff{r}=0$. At this stage we perform quantum
mechanical perturbation theory by expanding the denominator of the
Green function with respect to the higher order Coulomb potentials.
The third order corrections read
\begin{eqnarray}
&& \hspace{-.6cm} \delta_3\hat{G}=
-\,\hat{G}_0\delta V_3 \hat{G}_0 + \, 2\hat{G}_0\delta V_1
\hat{G}_0\delta V_2 \hat{G}_0
  \nonumber \\
 &&\hspace{-.0cm}
- \, \hat{G}_0\delta V_1\hat{G}_0\delta V_1\hat{G}_0\delta V_1
   \hat{G}_0 \, .
   \label{expandedGF}
\end{eqnarray}
Here $G_0=(\bff{p}^2/m_t-E)^{-1}$ is the zeroth order Green function,
$\delta V_n \equiv [{\cal V}_n\, V_C](r)$ is the n-th order Coulomb
potential. We calculated the expanded Green function
semi-analytically and obtained double sum representations, which were
evaluated numerically \cite{beneke-kiyo-schuller05}.

The Green function $G(E)$ has a single pole at the bound-state energy
level $E=E_n$,
\begin{equation}
G(E) \stackrel{E\to E_n}{=}\frac{|\psi_n(0)|^2}{E_n-E-i\epsilon}
+\mbox{non-singular} \, ,
\label{nearpole}
\end{equation}
where $|\psi_n(0)|^2$ and $E_n$ is the bound-state wave function
squared at the origin and energy level, respectively, which has 
series expansion in $\alpha_s$
\begin{eqnarray}
&&\hspace{-.7cm} E_n=E_n^{(0)}\, \big[1+\sum_{i=1}^3
\left(\frac{\alpha_s}{4\pi}\right)^i e_i \, \big]\, ,
\nonumber\\
&&\hspace{-.7cm} |\psi_n(0)|^2=|\psi_n^{(0)}(0)|^2\,
\big[1+\sum_{n}^3 \left(\frac{\alpha_s}{4\pi}\right)^i f_i \, \big]\,,
\label{energy_wf}
\end{eqnarray}
where $E_n^{(0)}, |\psi_n(0)|^2$ is the zeroth order result for the energy
and wave function. By performing matching between expanded Green function
eq.(\ref{expandedGF}) and the pole structure of exact Green function
eq.(\ref{nearpole}), we obtained analytical result 
\cite{beneke-kiyo-schuller05} for $e_i$ and $f_i$ for the S-wave 
bound state at NNNLO. The expression is too lengthy to show here. In the next section
we discuss phenomenological significance of the NNNLO Coulomb
corrections using the obtained result.

\section{Numerical analysis and Conclusion}

Our formalism up to now is not specific to the top quark, actually it
is applicable to bottom quark system replacing $m_t$ and $n_f=5$ by
$m_b$ and $n_f=4$ ($n_f$ exists in the $\alpha_s$ and
$\beta$-function in the Coulomb potential.) We investigate the
quarkonium energy levels and $t\bar{t}$ threshold cross section
in the following. 

In the phenomenological analysis
we use the potential-subtracted (PS) mass \cite{Beneke98} to make our
prediction infrared renormalon free. The relation between the pole and
PS masses is given by
\begin{eqnarray}
&&\hspace{-.6cm}
m = m_{\rm PS}(\mu_f) -\frac{1}{2}\int_{q \leq
\mu_f}
 \frac{d^3 {\bff{q}}}{(2\pi)^3}\,
 \tilde{V}(\bff{q})|_{\nu=\mu_f}\, ,
 \label{PSmass}
\end{eqnarray}
where $\mu_f$ is infrared cutoff of order $m \alpha_s$.
We take into account non-Coulomb corrections for quarkonium 
energy levels known from literatures
\cite{kniehl-penin99,kniehl-penin-smirnov-steinhauser00,penin-steinhauser02,penin-smirnov-steinhauser05}.
So the results are complete NNNLO as far as energy level is concerned.

\subsection{Bottomonium masses}
Using the analytical expression for the $e_i, f_i$, we obtain 
the relation between mass of bottomonium and bottom quark. It might be
instructive to show the results using pole and PS masses:
\begin{eqnarray}
&&\hspace{-.6cm } M_{\Upsilon(1S)}=2 m_b+E_1^{(0)}\big[ 1+
1.09_{\rm NLO}
\nonumber \\
&&\hspace{-.6cm}
+\big(1.42+0.36_{nC}\big)_{\rm N^2LO}
 +\big(2.29+0.28_{nC}\big)_{\rm N^3LO} \big]
\nonumber \\
&&\hspace{-.6cm} = 2 m_{b,\rm PS}+E_{1,\rm PS}^{(0)} \big[ 1+
0.19_{\rm NLO} \nonumber \\
&&\hspace{-.6cm} +\big(0.07-0.23_{nC}\big)_{\rm N^2LO}
+\big(0.09-0.19_{nC}\big)_{\rm N^3LO} \big], \nonumber
\end{eqnarray}
where $m_b=5$ GeV, $m_{\rm PS}(2GeV)=4.6$ GeV.
The numbers are given separately for Coulomb and
non-Coulomb to show numerical dominance of the former (in the pole
scheme). One can see the presence of anomalously large Coulomb
correction (IR renormalon) in the pole-mass scheme, while in the
PS mass scheme this behavior is improved and convergence of the
series became better.
\begin{figure}[htb]
\vspace*{-1cm}
\begin{center}
\makebox[0cm]{ \scalebox{0.55}
{\rotatebox{0}{
     \includegraphics{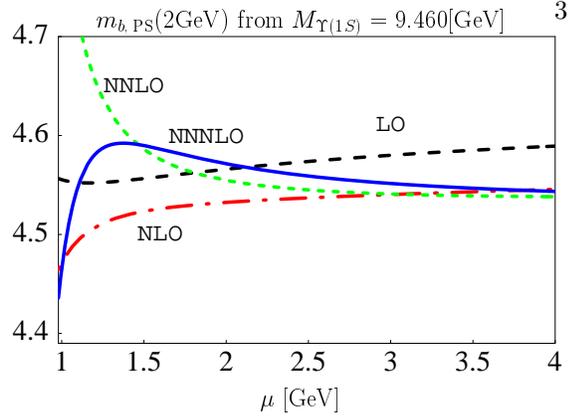}}}}
\end{center}
\vspace*{-1cm} \caption{ The bottom PS mass, $m_{b,\rm
PS}(2\,\mbox{GeV})$ extracted from the Upsilon mass}
\label{fig:PSmass}
\vspace*{-.5cm}
\end{figure}

We use the mass relation between $\Upsilon(1S)$ and $m_{b,{\rm PS}}$
to extract the bottom PS mass from the experimental value
$M_{\Upsilon(1S)}|_{exp.}=9.460$ GeV. We obtained
\begin{eqnarray}
&&\hspace{-.6cm}
m_{b,\rm PS}(2\,\mbox{GeV})
=(4.57 \pm 0.03_{\rm pert.}
\nonumber \\
&&\hspace{1cm} \pm 0.01_{\alpha_s} \pm 0.07_{\rm non-pert.}) \,\,
\mbox{GeV},
\label{finalres}
\end{eqnarray}
where the subscripts denote the source of errors.

\begin{figure}[htb]
\vspace*{-.5cm}
\begin{center}
\makebox[0cm]{ \scalebox{0.55}
{\rotatebox{0}{
     \includegraphics{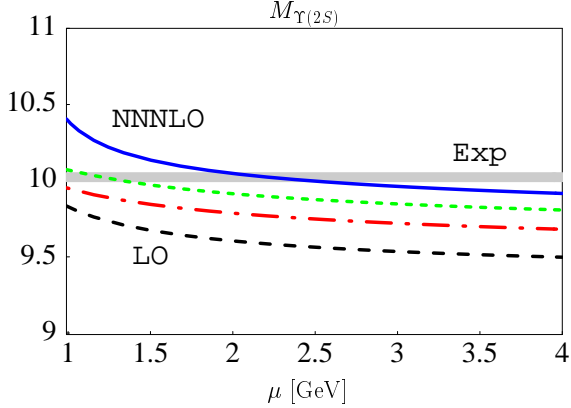}}}}
\end{center}
\vspace*{-1cm} \caption{$\Upsilon(2S)$ mass as a function of
  scale $\mu$. The lines are plotted for LO(long dashes),
NLO(long-short dashes),
  NNLO(dots), NNNLO(solid) and shaded band shows $M_{\Upsilon(2S)}|_{exp.}$.}
\label{fig:mb2S}
\vspace*{-.5cm}
\end{figure}

In Fig.\ref{fig:PSmass} we show a scale dependence of
the extracted PS mass $m_{b,{\rm PS}}(2{\rm GeV})$. 
Using extracted PS mass we are able to predict the masses of excited
states of the spin triplet S-wave $\Upsilon$ family. In Fig.\ref{fig:mb2S}
we show the $\Upsilon(2S)$ mass using $m_{b,PS}(2{\rm GeV})=4.57$ GeV
as a function of the renormalization scale $\mu$. One can see that
the large NNNLO corrections are preferable for scale $\mu > 2$ GeV,
however the prediction overshoots the experimental value 
$M_{\Upsilon(2S)}=10.023$ GeV at the lower scale. The naively expected
natural scale for the bottomonium is $\mu=C_F\alpha_s m_{b,\rm PS}/n$ 
($n$ is the principle quantum number), which is 
$\mu=1.23$ GeV for $\Upsilon(2S)$.
This may indicate break down of perturbative computation for the
excited $\Upsilon$ family. Similar behavior is observed for $\Upsilon(3S)$.

\subsection{Toponium}
In future linear colliders remnant of toponium $1S$ resonance
should be observed as a peak position of the $t\bar{t}$
cross section. This enables us to measure the $M_{t\bar{t}(1S)}$ 
and extract the top quark mass from the data.
Here we perform an exercise, how precisely we can predict the
1S toponium mass when we fix the top quark mass as an input parameter.
Adopting
$m_{t,\rm PS}=175\,$GeV and $\mu=32.6\,$ GeV 
($=C_F\alpha_s m_{t,\rm PS}$), we obtain
\begin{eqnarray}
&& \hspace{-.6cm} M_{t\bar t(1S)}=(350+0.85+0.05-0.13+0.01)\,\,{\rm
    GeV}
\nonumber \\
&&\hspace{.6cm} = 350.78\,\,\mbox{GeV}.
\end{eqnarray}
Scale variation between $15 < \mu < 60 {\rm GeV}$ changes the total
number only by $60$ MeV. The small higher order correction implies
that precise top quark mass extraction is possible in principle,
in the total cross section measurement. (There are several issues 
in the total cross section measurement at linear collider experiments, 
see for instance \cite{snowmass05}). To obtain \MS mass, being more
commonly used in high energy processes, from $M_{t\bar{t}(1S)}$ we
need to know the relation between the PS and \MS masses at 4-loop order,
which is currently unknown. Analysis using {\it large}-$\beta_0$
approximation to 4-loop \MS - pole mass relation and direct
extraction of \MS mass from $M_{t\bar{t}(1S)}$ is available from
ref.\cite{kiyo-sumino03}, which is consistent with our result.

\subsection{The Coulomb wave function at the origin and Green function}
In this subsection we discuss the Coulomb wave function and
the Green function. Since the complete NNNLO non-Coulomb
corrections and Wilson coefficient of the production current in EFT
are unknown, we shall discuss only the Coulomb corrections.
As we demonstrated in the previous section, applicability of this
method to bottomonium system is doubtful due to large NNNLO
correction and slow convergence. Thus we focus on the case of
toponium Coulomb wave function and Green function.

\begin{figure}[htb]
\vspace*{-.5cm}
\begin{center}
\makebox[0cm]{ \scalebox{0.5}{\rotatebox{0}{
     \includegraphics{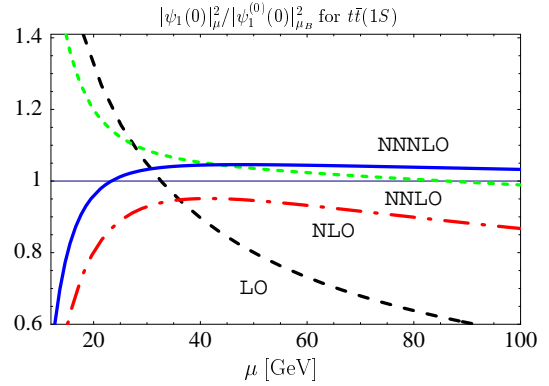}}}
}
\end{center}
\vspace*{-1.2cm} \caption{The Coulomb wave function at the origin
squared as a function of the scale $\mu$.} \label{fig:wf}
\vspace*{-.8cm}
\end{figure}

We show numerical formula for the $1S$ toponium Coulomb wave function
at the origin 
\begin{eqnarray}
&&\hspace{-.6cm}
\big|\psi_1(0)\big|^2_C
=
\frac{(m_t C_F\alpha_s)^3}{8\pi}
\big[
1
+\alpha_s \big(-0.4333
\nonumber \\
&&\hspace{-0.6cm}
+ 3.661\,L\big)
+\alpha_s^2 \big(5.832-5.112\,L + 8.933\,L^2\big)
\nonumber \\
&&\hspace{-.6cm}
+\,\alpha_s^3 \big(-13.73
                 +6.446\,\ln\big(\frac{\nu}{m_t C_F \alpha_s}\big)
                 + 39.72\,L
\nonumber \\
&&\hspace{-.6cm}
- 22.91\,L^2 + 18.17\,L^3\big)\,
\big]\, ,
\label{toppsi1}
\end{eqnarray}
where $L=\ln(\mu/C_F\alpha_s m_t)$. To draw Fig.\ref{fig:wf}
we rewrote the eq.(\ref{toppsi1}) using $m_{t,
  PS}(20 {\rm GeV})=175$ GeV and took into account the mass
correction eq.(\ref{PSmass})  at given order of
nonrelativistic expansion consistently. 
The figure shows that perturbative corrections
to the Coulomb wave function is reasonably small for $\mu > 25$ GeV
and the scale dependence is mild, while in lower
scale $\mu < 25$ GeV the corrections are too large so the
perturbative expansion is unreliable in the lower region.

\begin{figure}[htb]
\vspace*{-.5cm}
\begin{center}
\makebox[0cm]{ \scalebox{0.9}{\rotatebox{0}{
     \includegraphics{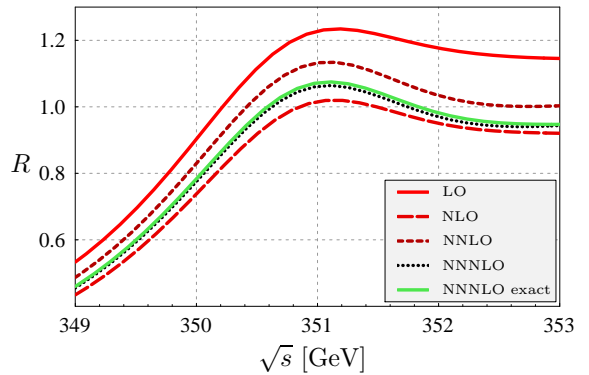}}}}
\end{center}
\vspace*{-1.3cm}
\caption{Top quark pair production (Coulomb
corrections only) for $\mu=30$ GeV, successively including higher
order corrections. } \label{fig:cs_order} \vspace*{-.7cm}
\end{figure}
In Fig.\ref{fig:cs_order} we show the $t\bar{t}$ threshold
cross section as a function of CM energy for
$m_{t,\rm PS} (20{\rm GeV})=175$ GeV including only the Coulomb
correction successively from LO to NNNLO. The line denoted
as ``NNNLO exact'' is a cross section obtained by numerically solving
shr\"oding equation for the Green function using NNNLO Coulomb potential.
The result show a convergence of the perturbative approximation
to the NNNLO-exact line.

\begin{figure}[htb]
\vspace*{-.5cm}
\begin{center}
\makebox[0cm]{ \scalebox{0.9}{\rotatebox{0}{
     \includegraphics{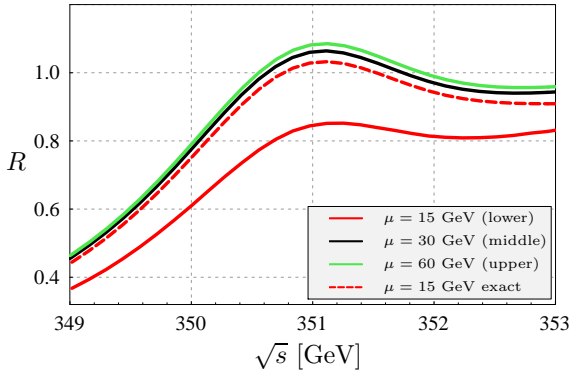}}}
}
\end{center}
\vspace*{-1.3cm} \caption{The NNNLO $t\bar{t}$ cross section (
Coulomb corrections only) for $\mu=15, 30, 60$ GeV for the
perturbative approximation, and $\mu=15$ GeV for the NNNLO exact
result.} \label{fig:cs_mu} \vspace*{-.7cm}
\end{figure}
In Fig.\ref{fig:cs_mu} we show the scale dependence of NNNLO cross
section for the perturbative Green function with $\mu=15,30, 60$ GeV, and
$\mu=15$ GeV for NNNLO-exact. The exact Green function is
stable against scale variation (so we showed only the case $\mu=15$ GeV),
while perturbative Green function (cross section) is unstable
against scale variation from 15 to 30 GeV (moderate change from $\mu=30$
to 60 GeV). This is consistent with wave function analysis, 
where the higher order corrections were
large for $\mu <25$ GeV, so we may conclude that the perturbative
expansion is not reliable in lower scale region. 
The NNNLO-exact contains higher order insertions of the Coulomb 
potential to all order in eq.(\ref{expandedGF}). 
The perturbative cross section agrees well with NNNLO-exact for
large scale where it is supposed to be reliable 
from the wave function analysis. We believe that the NNNLO-exact cross 
section is reliable in wider range of $\mu$, and the perturbative 
cross section is reliable only in the region $\mu > 25$ GeV. 
Indeed we find that the multiple insertions of the Coulomb 
potential give large contributions to the perturbative Green function, 
and is slowly converging at small scale. Thus we conclude that the 
``correct'' scale choice for the perturbative (Coulomb) cross section
is $ \mu>25$, while choice of small scale may lead to misleadingly
large uncertainties. We estimated yet unknown higher order Coulomb 
corrections should be less than 5 \%. 

\vskip0.4cm\noindent
{\bf Acknowledgements}\\
Y.K. would like to thank M. Beneke and K. Schuller for useful
discussion, and for reading this manuscript. Y.K. thanks 
J. Kodaira for invitation to the RADCOR05 conference.



\begin{thebibliography}{99}

\bibitem{beneke-kiyo-schuller05}
M. Beneke, Y. Kiyo and K. Schuler, Nucl. \ Phys. \ {\bf B714} (2005)
67.

\bibitem{Pineda-Soto97}
A.~Pineda and J.~Soto,
Nucl.\ Phys.\ Proc.\ Suppl.\  {\bf 64} (1998) 428 [hep-ph/9707481].

\bibitem{Luke-Manohar-Rothstein99}
M. E. Luke, A. V. Manohar and I. Z. Rothstein,
Phys.\ Rev.\ {\bf D 61} (2000) 074025. 

\bibitem{TopWGR}
A. H. Hoang et al., 
Eur.\ Phys.\ J. \ direct {\bf C 2} (2000) 1, and references therein.


\bibitem{kniehl-penin99}
B. A. Kniehl and A. A. Penin, Nucl.\ Phys.\ {\bf B 563} (1999) 200.

\bibitem{kniehl-penin-smirnov-steinhauser00}
B.A. Kniehl, A.A. Penin, V.A. Smirnov and M. Steinhauser,
Nucl.\ Phys.\ {\bf B 635} (2002) 357.

\bibitem{brambilla-pineda-soto-vairo99} 
N. Brambilla, A. Pineda, J. Soto and A. Vairo, Nucl.\ Phys.\ {\bf B
566} (2000) 275.

\bibitem{Beneke98}
M.~Beneke,
Phys.\ Lett.\ B {\bf 434} (1998) 115.

\bibitem{penin-steinhauser02}
A. A. Penin and M. Steinhauser, Phys. \ Lett.\ {\bf B 538} (2002)
335.

\bibitem{penin-smirnov-steinhauser05}
A. A. Penin, V. A. Smirnov and M. Steinhauser, Nucl. \ Phys. \ {\bf
B 716} (2005) 303.

\bibitem{snowmass05}
A. Juste et al., Report of the 2005 Snowmass Top/QCD Working Group [hep-ph/0601112].

\bibitem{kiyo-sumino03}
Y. Kiyo and Y. Sumino, Phys. \ Rev.\ {\bf D 67} (2003) 071501


\end{thebibliography}
\end{document}